\documentclass[preprint,amsmath,amssymb,showpacs]{revtex4}

\usepackage{graphicx} 
\usepackage{epsfig}

\begin{document}

\title{Poissonian bursts in e-mail correspondence }
\author{C. Anteneodo$^{a}$, R. Dean Malmgren$^{b}$, and D. R. Chialvo$^{c}$   }
\address{  (a)  Departamento de F\'{\i}sica, PUC-Rio and  
National Institute of Science and Technology for Complex Systems,    
 Rua Marqu\^es de S\~ao Vicente 225, CEP 22453-900 RJ, Rio de Janeiro, Brazil \\
 (b)   Department of Chemical and Biological Engineering, Northwestern University, 
 Evanston, IL 60208, USA \\
 (c)  Department of Physiology, Feinberg Medical School,
Northwestern University, 303 East Chicago Ave. Chicago, IL 60611, USA  
 }
\date{\today}
\begin{abstract}
Recent work has shown that the distribution of inter-event times for
e-mail communication exhibits a heavy tail which is statistically consistent with a cascading Poisson process.   
In this work we extend the analysis to higher-order statistics, using the Fano and Allan factors  
to quantify the extent to which the empirical data depart from
the known correlations of Poissonian statistics.
The analysis shows that the higher-order statistics from the empirical data is indistinguishable from
that of randomly reordered time series,  thus demonstrating that e-mail
correspondence is no more bursty or correlated than a Poisson process.  
Furthermore synthetic data sets generated by a cascading Poisson process 
 replicate  the burstiness and correlations observed in the empirical data.  
Finally, a simple rescaling analysis using the best-estimate rate of activity, confirms
that the empirically observed correlations arise from a non-homogeneus Poisson
process.
\end{abstract}


\maketitle 

The assessment of human activity patterns
is crucial for many applications, such as optimization of information traffic, service scheduling
and human resource planning.
In particular, the temporal dynamics of e-mail correspondence
sparked recent interest~\cite{eckmann,barabasi0,barabasi1,malmgren0,barabasi2,malmgren1,other}
 because of its importance as a communication medium and the availability of very large databases.
Recent research~\cite{barabasi0,barabasi1,zhou,malmgren0,malmgren1}
has shown that the probability distribution of the time elapsed between consecutively sent
e-mails by a single user exhibits heavy tails. 

The origin of such heavy-tailed statistics is controversial and the focus of much debate. One explanation for the existence of the inter e-mail times distribution is a priority queuing model \cite{barabasi0}. Another  contrasting view is 
 a cascading Poisson process~\cite{malmgren1,malmgren2}.  In
this latter model, there is a primary non-homogeneous Poisson process, which explicitly incorporates
daily and weekly modulations, each of whose events triggers a secondary process
which is also Poissonian but with a much larger characteristic rate.  According
to this model, ``bursts'' of e-mail activity occur in non-overlapping
homogeneous Poisson cascades (as opposed to the overlapping cascades of
Ref.~\cite{gruneis}, for instance) separated by long periods of inactivity
defined by the primary process.  The resulting inter-event time distribution
predicted by the model is therefore heavy-tailed due to the mixture of several
different scales of rates of activity.

While the cascading Poisson process has been shown to be statistically
consistent with empirical inter-event time distributions of several
individuals~\cite{malmgren1}, it is unclear whether higher-order statistical
patterns are present in the data and whether the cascading Poisson process
adequately captures these patterns. Here, we quantify the burstiness and correlations in the empirical and
synthetically generated point process data sets using standard statistical
measures.  We show that the burstiness and correlations in e-mail communication
patterns are Poissonian, which are in fact reproduced by the cascading Poisson
process.  
%%Figure 1%%%%%%%%%%%%%%%%%%%%%%%%%%%%%%%%
\begin{figure}[t!]
\centering 
\includegraphics*[bb=70 470 470 750, width=0.75\textwidth]{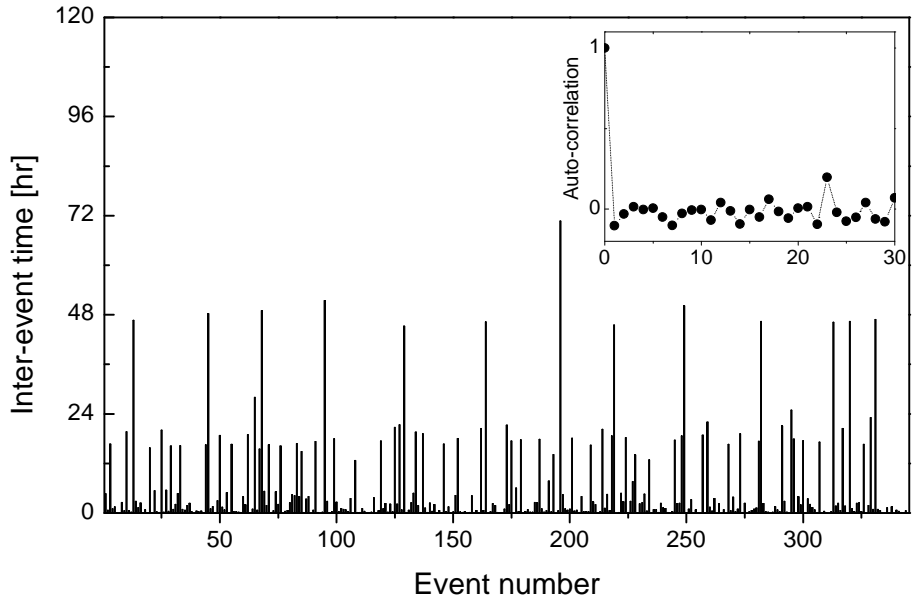}  
\caption{ 
Time series of inter e-mail intervals for a representative user, and its autocorrelation function (inset).
}
\label{fig:timeseries}
\end{figure}
%%%%%%%%%%%%%%%%%%%%%%%%%%%%%%%%%%%%%%%

We study here a database considered
previously~\cite{eckmann,barabasi0,barabasi1,malmgren0,malmgren1} comprised 
of 3,188 e-mail
accounts over an 83-day period at a European University.  From this
database, we restrict our analysis to the 394 ``typical'' users that send at
least 40 messages over 83 days~\cite{malmgren0,malmgren1}, referred here as 
``empirical'' time series. An example is presented in Figure 1 (for User 467) which depicts some
of the typical features, namely that there are long pauses between e-mails 
of the order of 16 and 48 hours and that the short intervals are uncorrelated as the inset shows.
The second set of time series to be analyzed, was generated by the cascading Poisson model in Ref~\cite{malmgren1}, one per user, referred to here as  ``synthetic'' time series.

For a  given user,  the sequence of e-mails in time can be seen as a
point process~\cite{cox}.  Typically, point processes
are characterized with inter-event times or counting statistics, so one might
naturally characterize higher-order statistical patterns of a point process
with multivariate distributions of these quantities.  Multivariate
distributions, however, are difficult to assess for time series with few
events, as in the present case.  We therefore use the Fano and Allan factors,
two standard metrics for point processes that provide reliable results for time
series with few events~\cite{cox,gruneis,thurner,ACh}, to gain insight into
the higher-order statistical structure of e-mail correspondence.

%%Figure 2%%%%%%%%%%%%%%%%%%%%%%%%%%%%%%
\begin{figure}[t!]
\centering 
\includegraphics*[bb=90 330 560 670, width=0.75\textwidth]{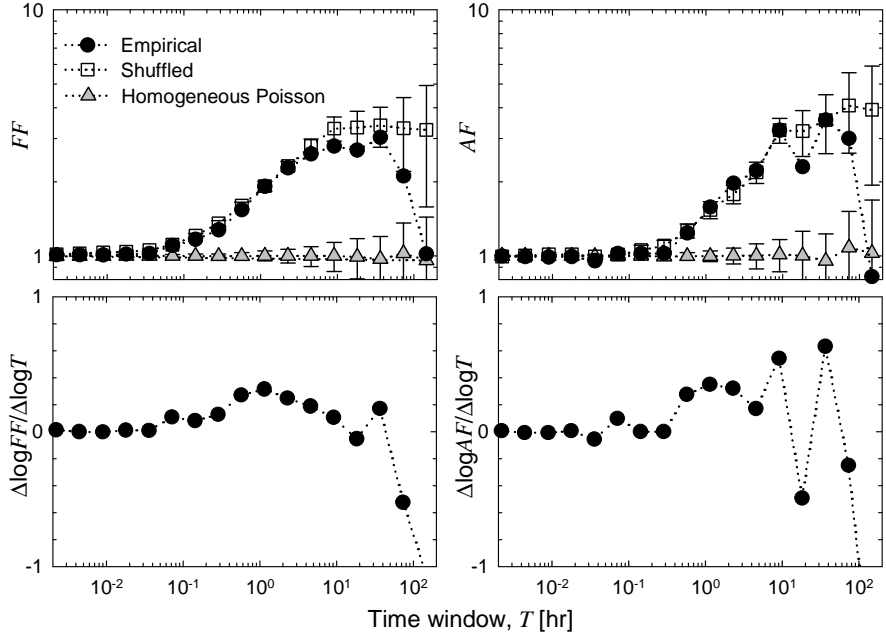}  
\caption{ 
Fano and Allan factors as a function of the time window $T$ for User 467, a
typical e-mail user (top panels).  For comparison, we plot the mean and
standard error of the Fano and Allan factors for 30 realizations of shuffled
time series as well as the mean and standard error of the Fano and Allan
factors for 30 realizations of the homogeneous Poisson process with the same
rate.  The bottom panels show the respective logarithmic local slopes of the
empirical Fano and Allan factor curves.
}
\label{fig:af_467}
\end{figure}
%%%%%%%%%%%%%%%%%%%%%%%%%%%%%%%%%%%%
The Fano and Allan factors are calculated by dividing the whole observation
time interval into $W$ non-overlapping time windows of equal length $T$ and
counting the number of events $N_k$ in each time window, indexed by $k$.
The Fano factor ($FF$) is the ratio of the variance to the mean of the number
of events in each time window, $FF=( \langle N_k^2\rangle-\langle
N_k\rangle^2)/\langle N_k\rangle$, and it represents a measure of the
dispersion---burstiness---of the resulting time series relative to a
homogeneous Poisson process with the same rate.  
The Allan factor ($AF$) quantifies the difference in variance of counts of adjacent time
windows,  $AF=(\langle (N_{k+1}-N_k)^2 \rangle)/(2\langle N_k\rangle), $ and it
is a measure of the correlation of counts between successive time windows
relative to the expectation from a homogeneous Poisson process with the same
rate.

If the time series were generated by a homogeneous Poisson process, then the
number of counts $N_k$ in each time window would be independent and identically
distributed random variables drawn from a Poisson distribution.  In such a case,
$FF(T)=AF(T)=1$, regardless of the time window length $T$ (Fig.~\ref{fig:af_467},
top panels, grey circles).
Deviations from unity therefore quantify departures from Poissonian statistics.
For example, $FF(T)>1$ would indicate that the time series is more bursty than
expected from a homogeneous Poisson process at a particular time-scale $T$.
Indeed, oftentimes researchers identify scale-free features in point processes
with a power-law increase in the Fano and Allan factors~\cite{thurner}.

%Figure 3%%%%%%%%%%%%%%%%%%%%%%%%%%%%
\begin{figure}[h]
\centering 
\includegraphics*[bb=160 290 490 630, width=0.75\textwidth]{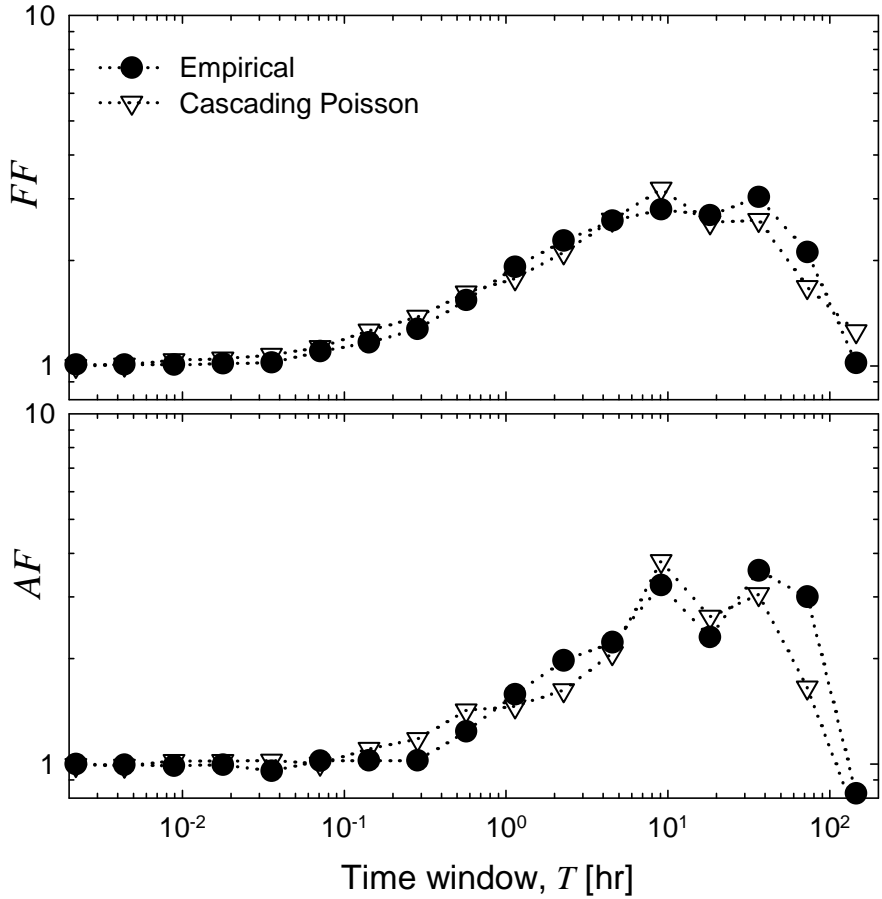}  
\caption{ 
Fano and Allan factors as a function of the length of the counting time window
$T$, for the empirical and synthetic time series for User 467.  The synthetic
time series is constructed with the same number of events as the empirical time
series from a cascading Poisson process with best-estimate parameters obtained
in Ref.~\cite{malmgren0}.
}
\label{fig:af_467syn}
\end{figure} 
%%%%%%%%%%%%%%%%%%%%%%%%%%%%%%%%%

We begin by analyzing the Fano and Allan factors as a function of the length of the
time window for a representative e-mail user, to be complemented later in the manuscript with
the averages for  the entire database.
The results for this user's activity are plotted in Fig.~{\ref{fig:af_467}}, 
(top panels, black circles, User 467 ). 
Notice that for time windows shorter than a few minutes 
the point process of e-mails  is essentially Poisson, denoted by the
fact that both
indices remain close to unity. 
For longer times the $FF$ and $AF$ curves noticeably depart 
from unity, which might suggest that there are some non-Poissonian 
effects in e-mail communication. Nevertheless, notice
that the non-unitary region exhibits  no power-law scaling.  
This can be also seen in the calculation of the local slope of the
log-log plot, which changes continuously denoting absence of scaling
(Fig.~{\ref{fig:af_467}}, lower panels). This finding is relevant at the light of
alternative models \cite{barabasi0} of this type of processes. 

To further evaluate the empirical time series' departure from
Poissonian statistics, we analyzed the surrogate time series obtained by
randomly reordering the sequence of inter-event times.  If the empirical time
series exhibits the same behavior in $FF$ and $AF$ as the shuffled time series,
then the observed departure from Poissonian statistics is only due to the
distribution of inter-event times and not due to their particular ordering; that is, the
inter-event times are independent, as is illustrated in Fig.~\ref{fig:af_467} (upper panel, white circles). 
Thus, the observed ``departure'' from Poissonian statistics seen here is 
merely an artifact of the heavy-tailed inter-event time distributions, and not of
some higher-order statistical structure.

Now, we proceed to compare the Fano and Allan factors of the empirical time series of
this typical user with a synthetic
time series generated from the cascading Poisson process from Ref.~\cite{malmgren1}. 
The model uses  the best-estimate parameters for this specific user 
and the same number of events as the empirical time series. 
As Fig.~\ref{fig:af_467syn} shows, 
the agreement between the empirical and synthetic $FF$ and $AF$ curves
is remarkable, indicating that the cascading Poisson process is capturing not only the
density distribution (already discussed at length in Ref.~\cite{malmgren1}) but also  
the higher-order statistical features of e-mail communication.  This
agreement is also consistent with (and anticipated by) the similarities 
 between the empirical and shuffled time series (Fig. ~\ref{fig:af_467}). 

%Figure 4%%%%%%%%%%%%%%%%%%%%%%%%%%
\begin{figure}[h!]
\centering 
\includegraphics*[bb=150 290 490 630, width=.75\textwidth]{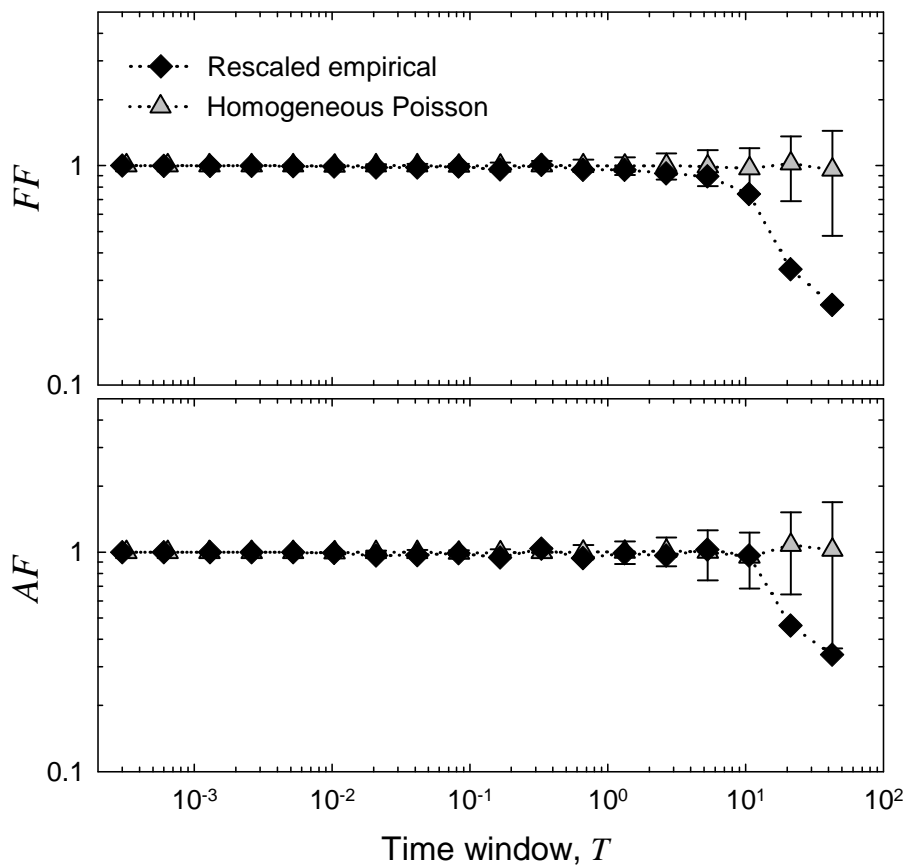}  
\caption{ 
Fano and Allan factors as a function of the time window $T$ for the rescaled
time series of User 467.  For comparison, we compute the Fano and Allan factors
for 30 realizations of a homogeneous Poisson process with unit rate and the
same number of events.  The large fluctuations observed at long time-scales are due to
poor statistics (e.g. $W=8$ in the longest time window).
}
\label{fig:af_467rsc}
\end{figure} 
%%%%%%%%%%%%%%%%%%%%%%%%%%%%%%%%%
The analysis up to now indicates that the origin of the observed
higher order correlations is related with the non-homogeneity in the
rate of e-mail activity. 
If this is the case, we should be able to rescale time such that the resulting point process
appears to have originated from a homogeneous Poisson process.
Specifically, any non-homogeneous Poisson process with occurrence rate
$\rho(t)$, can be mapped onto a homogeneous Poisson process through a simple
transformation of the timescale, namely $\tilde{t}=\int_0^t
\rho(t)$~\cite{cox}.  In this new time scale, the Poisson process has unit
rate, $\tilde{\rho}(\tilde{t})=1$.
In the particular case of a cascading Poisson process with known best-estimate 
parameters $\rho(t)$ and $\rho_a$  and where we know which events are associated 
with which process\cite{malmgren1}, we rescale the  times between consecutive
events accordingly.
The results presented in Fig.~\ref{fig:af_467rsc} confirm our hypothesis: the rescaled 
inter-event time sequence 
exhibit $FF$ and $AF$ values close to unity for time window lengths up to about ten
times the characteristic time. By comparison, the results corresponding to thirty realizations
of a homogeneous Poisson process with unit rate are presented as well.

%Figure 5%%%%%%%%%%%%%%%%%%%%%%%%%%%%%%
\begin{figure}[ht!]
\centering 
\includegraphics*[bb=85 325 550 660, width=0.75\textwidth]{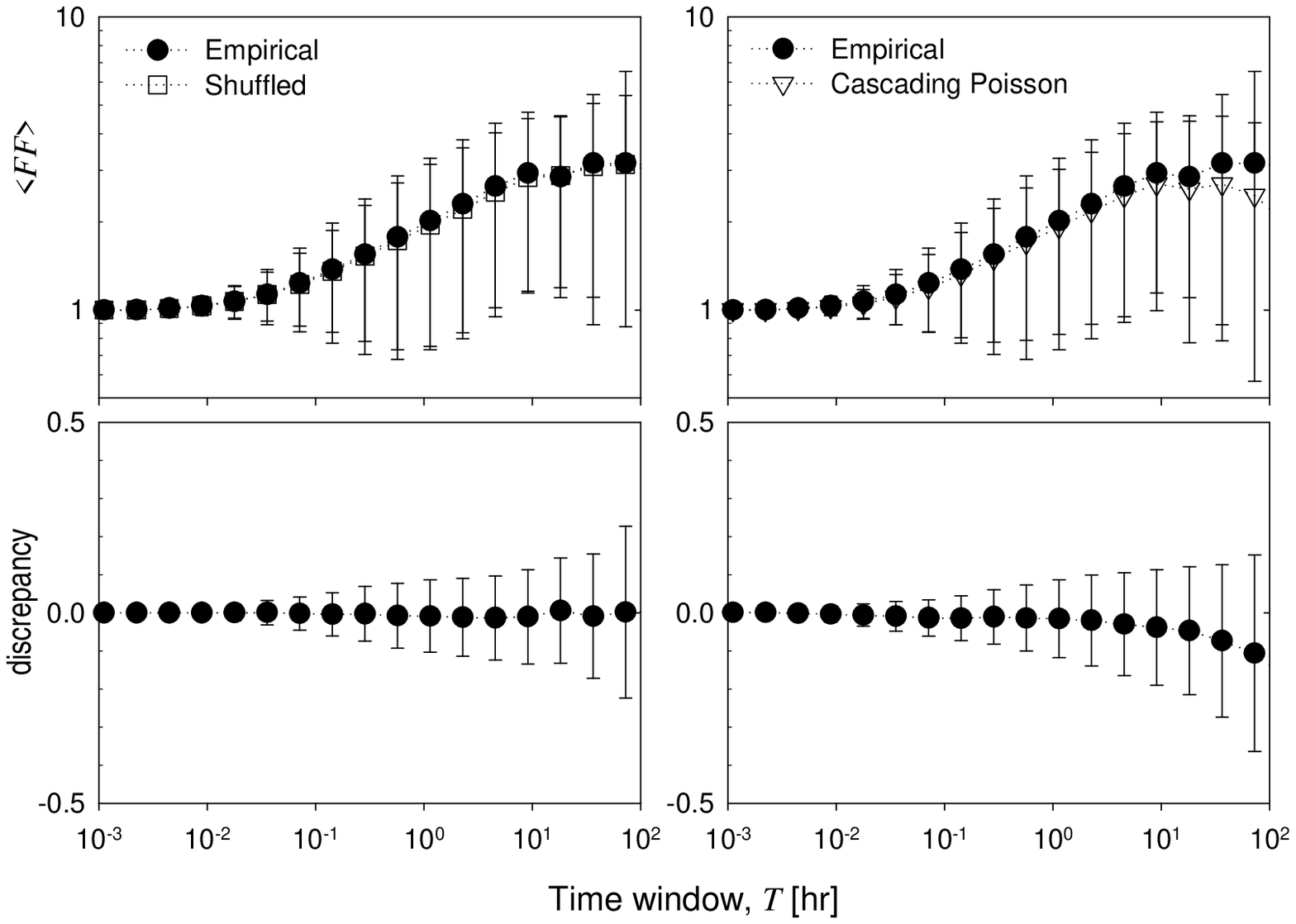}   
\caption{
Summary of agreement between the empirical time series, shuffled time
series (left panels) and cascading Poisson model (right panels).  
The top panels show  the mean and standard deviation (whiskers)
 Fano factor as a function of the time window length
for the time series averaged over all 394 users.
The bottom panels show the mean logarithmic distance
between  the $FF$ results for the
empirical and shuffled time series (left) or cascading Poissonian
 model (right) for all users.
}
\label{fig:allshuf}
\end{figure}
%%%%%%%%%%%%%%%%%%%%%%%%%%%%%%%%%%%

We know move beyond the analysis of User 467 to present the results
of our analysis of all 394 users. The results presented in Fig.~\ref{fig:allshuf} show the same
analysis presented in Figs. 2-4 now computed for all 394 e-mails users in the database.
In Fig.~\ref{fig:allshuf} (left top panel) it can be seen that the mean $FF$ curve
 for all users replicates well the mean $FF$ curve for the shuffled time series for each of the 394 users.
Thus, at the population level, higher order correlations of the process exhibit
the same basic features: for short time windows (up to a few
minutes), the $FF$ is unitary for all users, a fingerprint of a
Poissonian process; and, for longer time windows, both the empirical and
shuffled data sets exhibit a slight increase. As discussed before, this is indicative of
correlations in the empirical data which are  trivially related to the distribution of inter-event times. 
In that figure the mean logarithmic difference is also plotted, this is 
quantified, for each user and
time window size with log-distances,
$R_i(T)=\log_{10}[FF_{e,i}(T)/FF_{s,i}(T) ]$,where subscript $e$ indicates the empirical
time series and $s$ the series compared, either shuffled, synthetic or homogeneous Poisson. 
We do not find any significant deviations between the empirical and shuffled
$FF$ curves.
The analysis of the entire database also confirms the similarities 
between the empirical and synthetic  time series. This 
is presented in right top panel in Fig.~\ref{fig:allshuf}, demonstrating that
the mean behavior of the synthetic $FF$ curves is consistent with the empirical
mean $FF$ curve with no significant systematic deviations between them.

%Figure 6%%%%%%%%%%%%%%%%%%%%%%%%%%%%%%%
\begin{figure}[h!]
\centering 
 \includegraphics*[bb=120 210 540 700, width=0.75\textwidth]{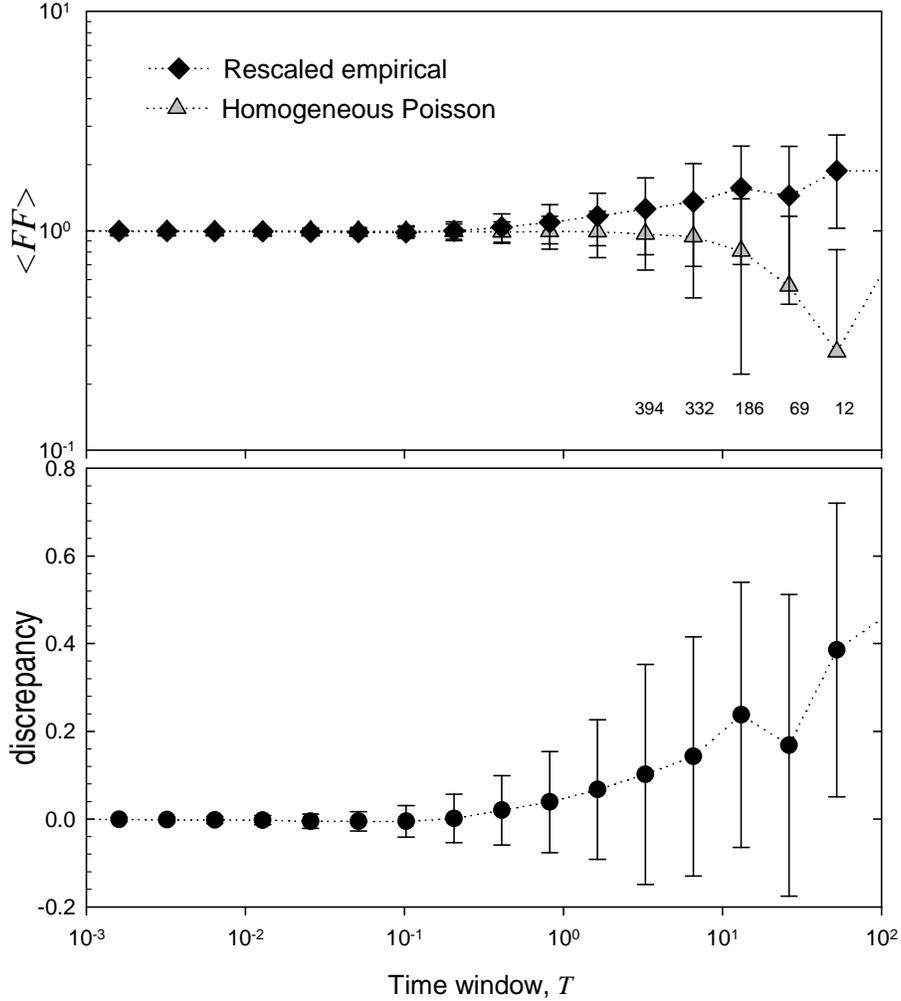}  
 \caption{
Top panel: Mean and standard deviation (whiskers) of
Fano factor as a function of the time window length
for the rescaled time series averaged over all 394 users.
Bottom panel:  shows the mean distance between the
$FF$ of the rescaled and that of a homogeneous Poisson process.}
\label{fig:allrsc}
\end{figure} 
%%%%%%%%%%%%%%%%%%%%%%%%%%%%%%%%%%%%%%
Similar conclusions can be reached from the results of the $FF$ curves 
of rescaled data compared to a homogeneous Poisson
process for all 394 users considered here. This is summarized in
Fig.~\ref{fig:allrsc}.   The statistical distance
indicates that after properly rescaling time,  
a homogeneous Poisson process is in fact  recovered.

Summarizing, the analysis described here shows that the 
sequences  of inter e-mail times are uncorrelated
for time scales shorter than a few minutes. At longer time 
scales, although the process exhibits  
correlations, they are trivial as they are found
to be originated by daily and weekly cycles of activity.
The trivial origin of these correlations is further confirmed by
a rescaling transformation which leads to  a homogeneous Poisson process. 
There was no evidence of scale free process at any of the levels
of e-mail activity analyzed.
To conclude, the present findings are consistent with the 
cascading non-homogeneous Poisson model as the best description
for the human activity patterns behind the e-mail sequences.

{\bf Acknowledgements:} We are grateful to J-P~Eckmann for providing the data.
CA acknowledges  Northwestern University for the kind hospitality and
Brazilian agencies CNPq and Faperj for partial financial
support, DRC acknowledges support by NIH NINDS of USA (Grants NS58661).

\end{document}